\documentclass[aps,prl,reprint,twocolumn,superscriptaddress,floatfix,nofootinbib]{revtex4-1}
\usepackage{amsmath,amssymb,color,comment}
\usepackage[makeroom]{cancel}
\usepackage[caption=false]{subfig}
\usepackage{mathrsfs}
\usepackage{graphicx}
\usepackage{subfig}
\usepackage[countmax]{subfloat}
\usepackage[english]{babel}
\definecolor{jadclr}{rgb}{0,0.5,0}
\definecolor{jadcolor}{rgb}{0.5,0,0}
\usepackage[bookmarks=true,colorlinks,linkcolor=jadclr,urlcolor=jadcolor,citecolor=jadcolor]{hyperref}
\DeclareMathOperator{\Tr}{Tr}

 \usepackage{dsfont}

\newcommand{\rmi}{{\rm i}}

\def\d{\mathrm d}

\definecolor{mypink1}{rgb}{0.858, 0.188, 0.478}

\def\d{\mathrm d}
\newcommand{\eh}[1]{\text{e}^{#1}}
\newcommand{\init}{_\text{i}}
\newcommand{\final}{_\text{f}}

\begin{document}

\title{Dynamical Quantum Phase Transitions: A Geometric Picture}

\author{Johannes Lang}
\affiliation{Physik Department, Technische Universit\"at M\"unchen, 85747 Garching, Germany}

\author{Bernhard Frank}
\affiliation{Physik Department, Technische Universit\"at M\"unchen, 85747 Garching, Germany}

\author{Jad C.~Halimeh}
\affiliation{Max Planck Institute for the Physics of Complex Systems, 01187 Dresden, Germany}
\affiliation{Physik Department, Technische Universit\"at M\"unchen, 85747 Garching, Germany}

\date{\today}

\begin{abstract}
The Loschmidt echo (LE) is a purely quantum-mechanical quantity whose determination for large quantum many-body systems requires an exceptionally precise knowledge of all eigenstates and eigenenergies. One might therefore be tempted to dismiss the applicability of any approximations to the underlying time evolution as hopeless. However, using the fully connected transverse-field Ising model (FC-TFIM) as an example, we show that this indeed is not the case, and that a simple semiclassical approximation to systems well described by mean-field theory (MFT) is in fact in good quantitative agreement with the exact quantum-mechanical calculation. Beyond the potential to capture the entire dynamical phase diagram of these models, the method presented here also allows for an intuitive geometric interpretation of the fidelity return rate at any temperature, thereby connecting the order parameter dynamics and the Loschmidt echo in a common framework. Videos of the post-quench dynamics provided in the supplemental material visualize this new point of view.
\end{abstract}

\maketitle

Equilibrium phase transitions are remarkable phenomena that have been under thorough experimental and theoretical investigation for decades. 
Over time, a number of advanced techniques such as scaling theory~\cite{Ma_book,Cardy_book,Sachdev_book} and the renormalization group method \cite{Wilson1971a,Wilson1971b,Wilson1972,Wilson1974,Fisher1974,Wilson1975} have been developed for the determination of the universal properties close to a critical point. 
One might ask whether an in-depth study of dynamical critical phenomena far from equilibrium is possible along the lines established in the equilibrium framework.
With the advent of modern ultracold atom \cite{Levin_book,Yukalov2011,Bloch2008,Greiner2002} and ion-trap \cite{Porras2004,Kim2009,Jurcevic2014} experiments, this originally purely academic question has become accessible in laboratories as well. 

Dynamical quantum phase transitions (DPTs) occur in the dynamics of a quantum system after quenching a set of control parameters $\{\Gamma\}$ of its Hamiltonian: $H(\{\Gamma\init\}) \to H(\{\Gamma\final\})$. Recently, the study of DPTs has focused on two largely independent concepts~\cite{Zvyagin2016}.
The first one, DPT-I~\cite{Moeckel2008,Moeckel2010,Sciolla2010,Sciolla2011,Gambassi2011,Sciolla2013,
Maraga2013,Chandran2013,Smacchia2015,Halimeh2017b,Mori2017,Zhang2017}, resembles equilibrium Landau theory: A system undergoes a dynamical phase transition if the long-time limit of the order parameter is finite for one set $\{\Gamma\init,\Gamma\final\}$, whereas it vanishes for different final parameters $\{\Gamma\final\}$. Furthermore, DPT-I also entails criticality in the transient dynamics of the order parameter and two-point correlators before reaching the steady state, giving rise to effects such as dynamic scaling and aging, which have been investigated theoretically \cite{Chiocchetta2015, Marcuzzi2016, Chiocchetta2017} and also observed experimentally \cite{Nicklas2015}.

The second concept, DPT-II, generalizes the non-analytic behavior of the free energy at a phase transition in the thermodynamic limit (TL) to the out-of-equilibrium case. To this end, the LE has been introduced as a dynamical analog of a free energy per particle \cite{Heyl2013}. DPT-II has been extensively studied both theoretically \cite{Heyl2013,Heyl2014,Andraschko2014,Vajna2014,Heyl2015,Vajna2015,
Budich2016,Bhattacharya2017,Heyl2017,Heyl_review} and in experiments \cite{Flaeschner2018,Jurcevic2017}. As we aim to calculate dynamical phase transitions at finite preparation temperatures, we define the distance covered in Hilbert space between the pre-quench density matrix $\rho\init$ and the time-evolved $\rho(t)=\exp (-\text{i} H\final t)\rho_\text{i}\exp (\text{i} H\final t)$ in the limit of infinite system size $N$ as the fidelity return rate \cite{Sedlmayr2018,Mera2018,Zanardi2007,Venuti2011}
\begin{align}
r_\text{F}(t)&=- \lim_{N \to \infty} \frac{1}{N} \ln\left|\Tr \sqrt{\sqrt{\rho \init} \rho(t) \sqrt{\rho\init}}\right|^2\,.
\label{eq:rfiniteT}
\end{align}
For $T \to 0$, the return rate reduces to the original definition $r(t)= - \lim_{N\to \infty}N^{-1}\ln|\langle\Psi_0| \eh{-\text{i} H\final t} |\Psi_0  \rangle|^2$~\cite{Heyl2013}. Within our semiclassical theory we will find a simple and intuitive expression for the distance measure~\eqref{eq:rfiniteT}.
  
DPT-II is characterized by cusps occurring in $r_\text{F}(t)$ at critical times $t_\text{c}$ after a quench. There are two scenarios for these cusps. The first is when the argument of the logarithm is zero, which is encountered only at $T=0$~\cite{Sedlmayr2018} in two-band models of free fermions \cite{Huang2016,Dutta2017,Budich2016}, when for critical quasi-momenta $\mathbf{k}_\text{c}$ the population becomes inverted \cite{Vajna2015}. Upon integration over $\mathbf{k}$-space the resulting logarithmic divergence will be turned into a cusp. Alternatively, the argument of the logarithm may itself become nonanalytic \cite{Halimeh2017,Zauner2017}, which occurs for example in nonintegrable quantum Ising chains with ferromagnetic power-law interactions \cite{Dutta2001,Knap2013,Jaschke2017,Vanderstraeten2018}. At $T=0$, numerical investigations have shown a relationship between the DPT-I and DPT-II in the presence of sufficiently long-range interactions \cite{Halimeh2017,Homrighausen2017,Zauner2017,Zunkovic2018}. At finite temperatures the DPT-I and DPT-II phase diagram based on $r_\text{F}$ coincide for the FC-TFIM~\cite{Lang2017}.

Of course, like for equilibrium phase transitions, a perfectly sharp cusp of any LE will only be observable in the TL. An accurate determination of the LE in this limit, however, requires computation of overlaps between different eigenstates to a precision that grows exponentially with system size. On the one hand, this sensitive dependence on $N$ complicates numerical treatment of large systems necessary for a reliable finite-size scaling, and, on the other hand, it may seem to completely rule out any kind of perturbative expansion with algebraic corrections. Here we show otherwise.

As we will discuss in detail in the case of the FC-TFIM, for models where MFT can be applied, one can create a controlled, semiclassical extension to the solution of the mean-field equations, which accurately reproduces the full return function and in particular determines the DPT-II phases correctly. In fact, a closely related analysis has already successfully explained the collapse and revival of the time-of-flight interference patterns following a quench to the deep lattice limit of the Bose-Hubbard model~\cite{Greiner2002}.
The Hamiltonian of the FC-TFIM, also known as the LMG model in nuclear physics \cite{Lipkin1965,Meshkov1965,Glick1965,Botet1982,Botet1983,Ribeiro2008}, reads
\begin{align}
H=-\frac{J}{8 N} \sum_{i\neq j}{\sigma_i^z \sigma_j^z}-\frac{\Gamma}{2} \sum_{i} \sigma_i^x\;,
\label{eq:Hstart}
\end{align}
where $\sigma^{\{x,y,z\}}_i$ are the Pauli matrices on site $i$. The normalization ensures extensive scaling of the energy. Furthermore, we set the ferromagnetic coupling $J$ to unity.
In order to study the DPT-II induced by quenches in the transverse-field strength $\Gamma$, we utilize the infinite-range interaction to rewrite the Hamiltonian in terms of the total spin $S_{\{x,y,z\}}=\sum_i \sigma_i^{\{x,y,z\}}/2$:
\begin{align}
H=-\frac{1}{2 N} S_z S_z- \Gamma S_x\;,
\label{eq:Hspin}
\end{align}
which is exact up to an irrelevant constant.
Due to $[H, \mathbf{S}^2]=0$, the total spin length $S$ is conserved, even after the quench $\Gamma\init\to\Gamma\final$.

For this quench protocol, the DPT-I phase diagram for $m_z=\langle S_z \rangle/N $ is completely determined by MFT \cite{Sciolla2011}, which is equivalent to the leading order of a $1/N$-expansion. It is based on the Bloch sphere representation of the spin in terms of the continuous classical vector $\mathbf{S}= S(\sin \theta \cos \phi, \sin \theta \sin \phi, \cos \theta)$ that contributes the highest weight to the free energy arising from the pre-quench Hamilton function
\begin{align}
H_\text{i}(\theta,\phi) = -\frac{S^2}{2N}\cos^2 \theta - \Gamma\init S\sin\theta\cos\phi\;.
\label{eq:Hclassical}
\end{align}
The short-time evolution is then governed by the classical equations of motion (EOM) derived from the post-quench Hamiltonian $H\final(\theta, \phi)$; see~\eqref{eq:eom} below. The MFT thus forms the starting point for the semiclassical treatment of the LE and the DPT-II phase diagram, the construction of which we will now detail. For simplicity, we restrict ourselves to $\Gamma\init=0$ in the rest of the manuscript. Initially, we focus on the zero-temperature case and deal with thermal states later. 

At $T=0$, one first finds the vector $\mathbf S_\text{cl}$ minimizing $H\init(\theta, \phi)$, which we choose, due to the spontaneously broken $\mathbb{Z}_2$ symmetry, to be fully polarized along the positive $z$-axis. In other words, $\mathbf{S}_\text{cl}$ has angular variables $\theta_\text{cl}=0$, $\phi_\text{cl}$ arbitrary, and the maximal possible length $S_\text{cl}=N/2$. For later convenience we also introduce $s=2S/N \in [0,1]$, so here $s_\text{cl}=1$.   

Next we have to quantize our theory in order to define the notion of overlaps between different states, which inherently arises from quantum mechanics. To do so, we assign to $\mathbf S_\text{cl}$ the spin WKB wave function \cite{vanHemmen1986,Braun1993,vanHemmen2003} of a quantum mechanical degree of freedom in the ground state of the energy landscape $H\init(\theta, \phi)$:
\begin{align}
\begin{split}
\Psi_0\left(\theta\right)=\mathcal{N} \text{e}^{-\frac{1}{2}S_\text{cl}\sin^2{\theta}}\;,
\label{eq:WKB}
\end{split}
\end{align}
where $\mathcal{N}\sim\sqrt{S_\text{cl}}$ is an inconsequential normalization; see SM. To enforce $\mathbb{Z}_2$ symmetry breaking, we restrict $\theta$ to the northern hemisphere. By construction $\Psi_0(\theta,\phi)$ correctly determines the fluctuations $\langle \hat{S}_x^2\rangle=S_\text{cl}/2=\langle \hat{S}_y^2\rangle$ and $\langle \hat{S}_z^2\rangle=S_\text{cl}^2$ with next-to-leading order corrections in $N$.
\begin{figure}
\centering
\includegraphics[width=.85\columnwidth]{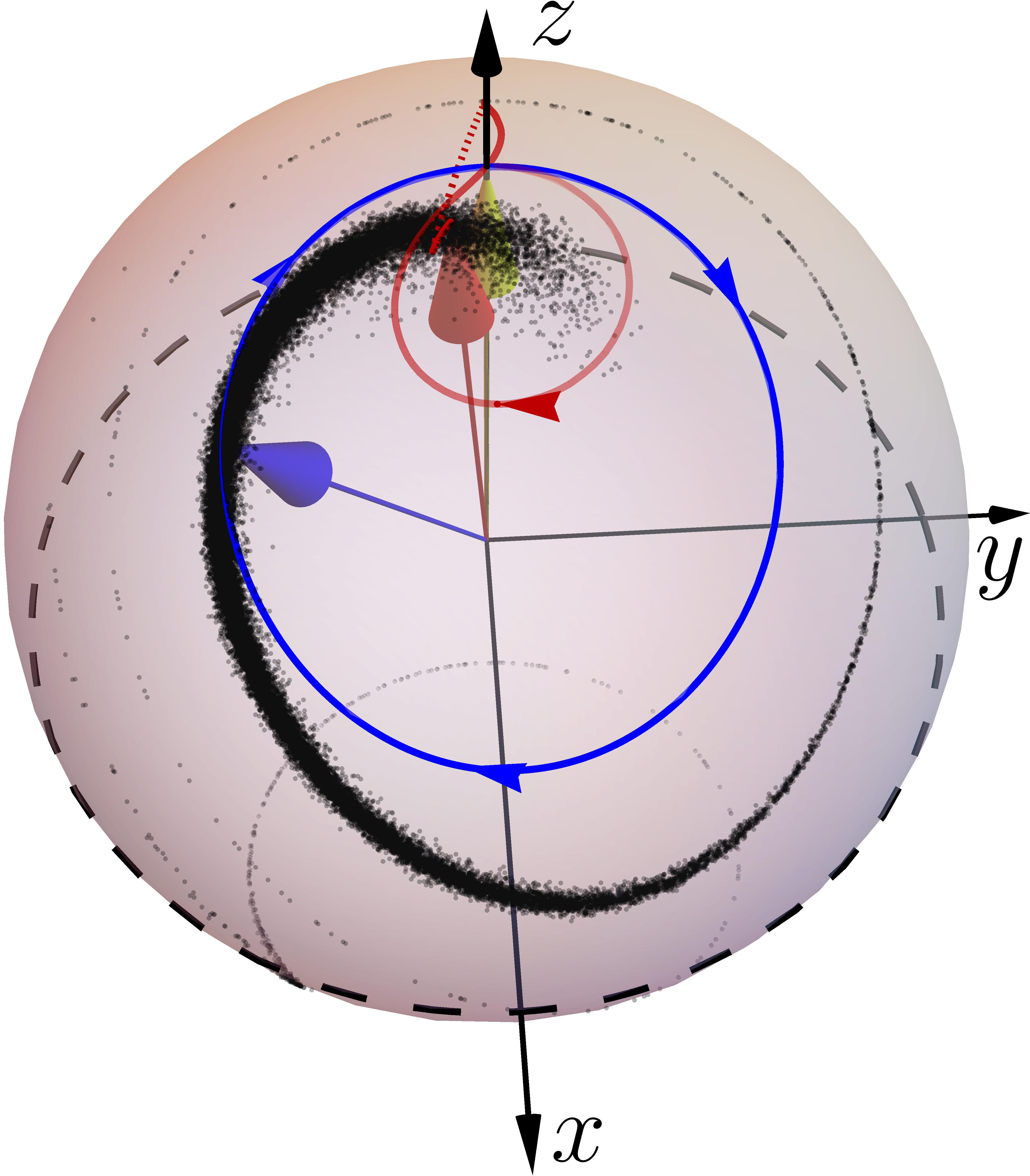}
\caption{(Color online) Semiclassical representation of the return rate on the Bloch sphere of the zero-temperature anomalous quench $\Gamma\init=0\to\Gamma\final=0.2$ shortly after the first critical time. The initial state pointing to the north pole is depicted by a yellow vector. The time-evolved classical initial state (blue vector) 
$\mathbf{v}_{\text{cl}}(t)=(\vartheta_\text{cl}(0, \phi \,|t),\varphi_\text{cl}(0,\phi\,|t))$
that governs the dynamics of the magnetization order parameter and thus determines the DPT-I phase, moves along the blue trajectory. The cloud of black dots indicates the distribution of the wave function that initially was centered symmetrically around the north pole. Finally the red arrow, which follows the red line, marks the orientation of the Loschmidt vector $\mathbf{v}_\text{max}(t)=(\vartheta(\bar{\theta}, \bar{\phi} \,|t),\varphi(\bar{\theta}, \bar{\phi} \, | t))$. At the critical time the sudden jump (dashed red line) of this saddle point orientation from the trailing to the leading edge of the time-evolved quantum amplitude results in a cusp in the return rate.}
\label{fig:anomalous} 
\end{figure}
\begin{figure}
\centering
\includegraphics[width=.85\columnwidth]{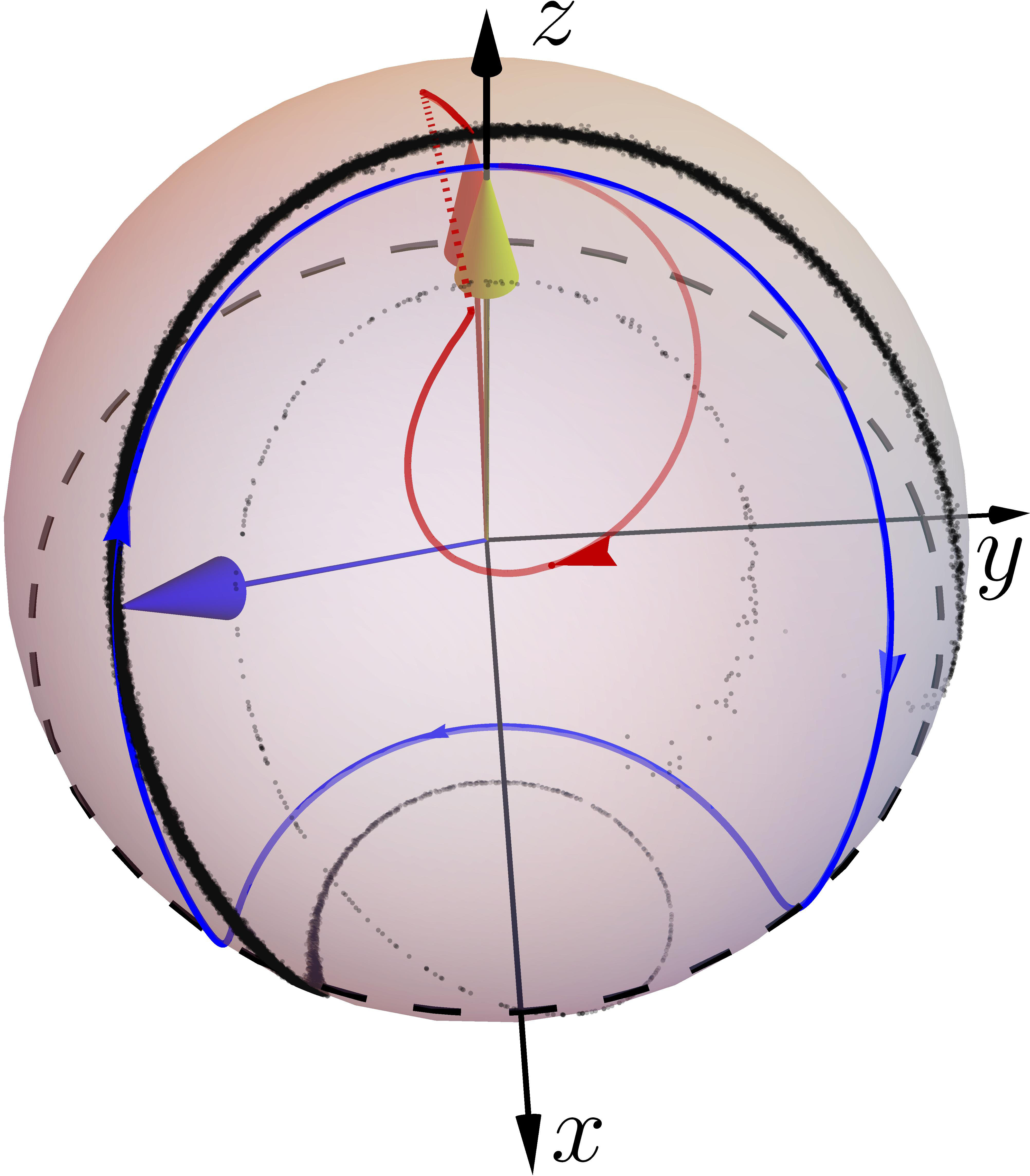}
\caption{(Color online) Depiction of the semiclassical spin configuration of the regular quench $\Gamma\init=0\to\Gamma\final=0.3$ at $T=0$ shortly after the first cusp. The color coding is the same as in Fig.~\ref{fig:anomalous}.}
\label{fig:regular} 
\end{figure}
\begin{figure}
\centering
\includegraphics[width=.85\columnwidth]{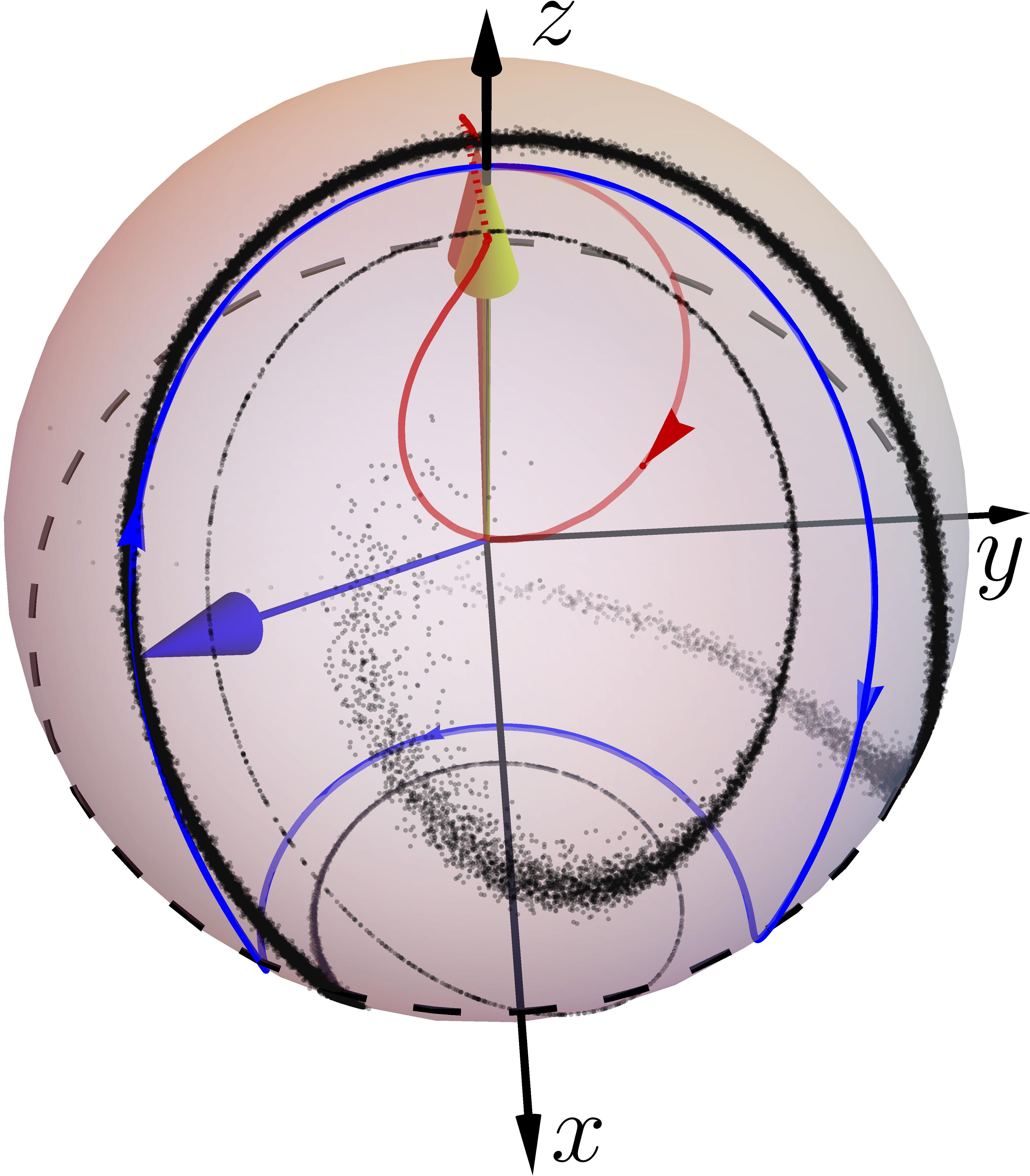}
\caption{(Color online) Illustration of the Bloch sphere in case of the regular quench $\Gamma\init=0\to\Gamma\final=0.2$ at the finite inverse temperature $\beta=5$. The color coding is the same as in Figs.~\ref{fig:anomalous} and \ref{fig:regular}.}
\label{fig:thermal} 
\end{figure}
\begin{figure}
\centering
\includegraphics[width=.95\columnwidth]{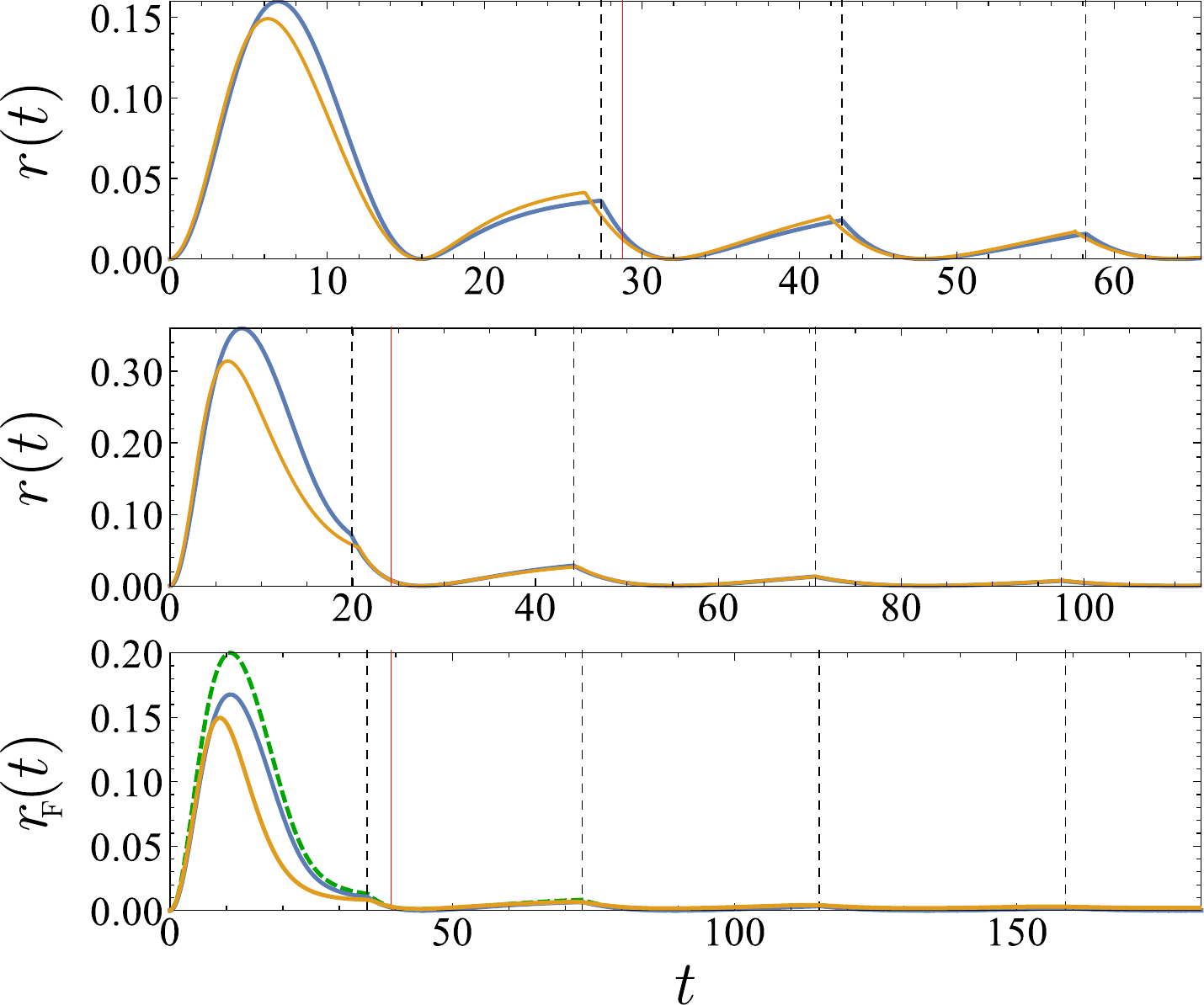}
\caption{(Color online) Comparison between the quantum return rate calculated numerically for a system of size $N=4001$ in yellow and the semiclassical results obtained from~\eqref{eq:rsemicl} and~\eqref{eq:ourrF} in blue. Panels a) through c) correspond to the parameters in Figs.~\ref{fig:anomalous} through \ref{fig:thermal} in that order. The red line indicates the time, depicted in those figures, while the black dashed lines mark the positions of cusps in the semiclassical return rate. The dashed green line in Panel~c) was calculated using the purely classical return rate from~\eqref{eq:rsimp}.}
\label{fig:LE_comp} 
\end{figure}

Having set up the semiclassical state at time $t=0$, we now incorporate the time evolution with $H\final$ by first determining the classical trajectories of the angular variables $(\vartheta(\theta,\phi \,|t),\varphi(\theta,\phi\, |t))$, which result from the classical EOM
 \begin{align}
\frac{\d \vartheta}{\d t}=\Gamma\final \sin\varphi,\qquad\frac{\d \varphi}{\d t}= \Gamma\final \cot\vartheta \cos\varphi\;,
\label{eq:eom}
\end{align} 
with initial conditions $(\vartheta(\theta,\phi|0),\varphi(\theta,\phi|0))=(\theta,\phi)$.
These derive from the Heisenberg equations for the total spin operators $S_{\{x,y,z\}}$ by neglecting all commutators that are suppressed by at least $1/N$ \cite{Lang2017}.
In close analogy to the time evolution in a truncated Wigner approximation \cite{Polkovnikov_review}, the initial amplitude $\Psi_0(\theta)$ is then transported along the classical trajectory, which implies that $\Psi_0(\vartheta(\theta,\phi\,|t))$ depends on both initial angles $\theta$ and $\phi$. Due to the absence of any dephasing within this description the magnetization, however, will never relax. Higher-order corrections can be treated by more faithfully representing the Schr\"odinger equation on the Bloch sphere, which will then include derivatives acting on the wave function~\eqref{eq:WKB}~\cite{Sciolla2011}. Here we take no effects beyond~\eqref{eq:eom} into account, which will turn out to determine the critical times accurately. In this limit the Loschmidt return function at $T=0$, defined in~\eqref{eq:rfiniteT}, reads   
\begin{align}
\begin{split}
r(t)&=-\frac{1}{N} \ln\left|\int \!\! \d\Omega\; \Psi_0^\ast
\!\big(\vartheta(\theta,\phi\,|t)\big) \Psi_0\big(\theta\big)\right|^2 \\
&= \frac{1}{2}\left(\sin^2{\vartheta(\bar\theta, \bar\phi\, |t)}+\sin^2{\bar{\theta}}\right)+\mathcal{O}\left(N^{-1}\right)\;,
\end{split}
\label{eq:rsemicl}
\end{align}
where the integral sums over the surface of the Bloch sphere with measure $\d\Omega=\d\phi\; \d \theta \sin\theta$.
The simple expression in the second line results from the limit $N\to \infty$, where, due to the extensive scaling of the exponent of the wave function~\eqref{eq:WKB}, at every moment in time the integral is determined by the Loschmidt vector $\mathbf{v}_\text{max}(t)=(\vartheta(\bar{\theta}, \bar{\phi} \,|t),\varphi(\bar{\theta}, \bar{\phi} \, | t))$ corresponding to the saddle-point trajectory that minimizes the exponent. Note that, as depicted in the SM, the initial coordinates $(\bar \theta, \bar \phi)$ are themselves time dependent. Furthermore, this result allows for a simple geometric interpretation: The classical trajectory with smallest arithmetic mean of initial and time-evolved WKB distances $A_0=-2\Re\ln{(\Psi_0/\mathcal{N})}/N=\sin^2{\theta}/2$ from the classical initial state dominates the LE. 

To compute $r(t)$ according to~\eqref{eq:rsemicl}, we cover the Bloch sphere with a Fibonacci lattice, assigning to each point the corresponding WKB amplitude of~\eqref{eq:WKB}. This lattice is then evolved in time, by numerically solving~\eqref{eq:eom} and finally extracting the site that yields the largest contribution to $r(t)$.  

Figures \ref{fig:anomalous} and \ref{fig:regular} illustrate our results for the spin dynamics in case of quenches to $\Gamma\final=0.2$ and $\Gamma\final=0.3$ shortly after the first critical time. The corresponding return rates can be found in Figs.~\ref{fig:LE_comp} a) and b). 
Movies of the spin dynamics are attached as video supplemental \cite{video_anomalous,video_regular,video_thermal}. The first quench is known to lie within the anomalous phase (no cusp in the first period(s) of $r(t)$) whereas the latter gives rise to a regular signal (all periods show non-analyticities) \cite{Halimeh2017}.

In the anomalous quench to $\Gamma\final=0.2$ \cite{video_anomalous}, the classical state $\mathbf{v}_{\text{cl}}(t)=(\vartheta_\text{cl}(0, \phi \,|t),\varphi_\text{cl}(0,\phi\,|t))$ moves only in the upper hemisphere yielding a positive $m_z$ at all times. Consequently, its trajectory returns so quickly to the initial state that the wave packet remains sufficiently concentrated around the classical state to prevent any discontinuous movement of the Loschmidt vector $\mathbf{v}_\text{max}(t)$ (obtained from \eqref{eq:rsemicl}) during the first period. The first jump of $\mathbf{v}_\text{max}(t)$, and therefore cusp in $r(t)$, appears only in the second period in agreement with the results obtained by ED calculations (see Fig.~\ref{fig:LE_comp}). At very late times the initial wave packet has spread so far over the Bloch sphere that the Loschmidt vector always points near the north pole, resulting in a very small $r(t)$.

For the regular quench to $\Gamma\final=0.3$ on the other hand \cite{video_regular}, the classical vector crosses the equator of the Bloch sphere where the increased fluctuations in $S_z$ result in a fast squeezing of the wave packet. This gives rise to a jump of the dominant orientation already during the first period of the motion, and thus to a regular LE.

The semiclassical evolution, therefore, allows for a very intuitive understanding of the relation between the order parameter dynamics and the return rate.

Let us now consider finite temperatures where the initial classical state for $\Gamma\init=0$ minimizes the mean-field free energy 
\begin{align}
F= -T \ln \left[\int \d\Omega \int_0^1 \d s\; s^2 D(Ns/2) \eh{-\beta H\init\left(\theta\right)} \right]\;,
\label{eq:MFFreeEnergy}
\end{align} 
where
\begin{align}
D(S)=\binom {N}{N/2-S}-\binom{N}{N/2-S-1}
\label{eq:degen}
\end{align}
denotes the degeneracy of the spin subspace of length $S$. These two equations specify the mean-field pre-quench state in terms of $S_\text{cl}$ with $\theta_\text{cl}=0$ and $\phi_\text{cl}$ arbitrary~\cite{Lang2017}. 
The exact initial density matrix $\rho \init = Z^{-1}\sum_n  \exp{(-\beta E_n)} |E_n \rangle\langle E_n| $ in the eigenbasis $|E_n\rangle$ of $H\init$ in our semiclassical description becomes
\begin{align}
\!\!\rho\init(\theta,\theta')\!=\!\frac{S_\text{cl}^2}{Z}\!\!\!\int \!\!\d\Omega\!\! \int_{-1}^1\!\!\!\!\!\!\d\!\cos{\theta_n}\Psi^*\!(\theta,\!E_n)\Psi(\theta'\!,\!E_n)\text{e}^{-\beta E_n}\,,
\label{eq:rho}
\end{align}
where the generalization of the WKB wave function in \eqref{eq:WKB} to an arbitrary eigenenergy $E_n=-s_\text{cl}S_\text{cl}\cos^2{\left(\theta_n\right)}/4$ reads
\begin{align}
\Psi(\theta,E_n)=\mathcal{N} \exp{\left[\frac{E_n b^2}{d^{3/2}}\big(\sinh{\left(4y\right)}-4y\big)\right]}.
\label{eq:WF}
\end{align}
Here, $\mathcal N$ is an inconsequential static normalization and $d=24 E_n/N+4s_\text{cl}^2$, $b=\sin\theta_n$, and $y=\text{arcsinh}\sqrt{Nd\left(b-\sin{\theta}\right)/\left(16 bE_n\right)}$ are functions of $\theta$ and $\theta_n$, where the latter seperates the classically allowed ($\theta<\theta_n$) from the forbidden region (see SM for the derivation).
In the TL the off-diagonal terms in \eqref{eq:rho} are suppressed by factors exponentially large in the system size and thus we can set $\theta'=\theta$.

Using this diagonal form of $\rho\init$ and the fact that the truncated time evolution acts only on the coordinates $(\vartheta(\theta, \phi \, |t),\varphi(\theta, \phi \, | t))$, we can write for the fidelity LE $r_\text{F}(t)= -\lim_{N\to \infty}N^{-1} \ln|\Tr \sqrt{\rho(0)} \sqrt{\rho(t)}|^2$; cf.~\eqref{eq:rfiniteT}. In the TL the remaining integrals in this expression once again reduce to their saddle-point values, equivalent to the minimization problem over all starting points $(\theta,\phi)$ in
\begin{align}
r_\text{F}(t)=&\min_{(\theta, \phi)}\bigg\{\text{dist}(\vartheta(\theta, \phi \, |t))+\text{dist}(\theta)\bigg\}\;
\label{eq:ourrF}
\end{align}
and all classical angles $\theta_n$ in the combined thermal and WKB distance measure
\begin{align}
\text{dist}(\theta)=\min_{\theta_n}\left\{\frac{\beta s_\text{cl}^2}{8}\sin^2{\theta_n}+\Re A\left(\theta,\theta_n\right)\right\}.
\end{align}
The geometric interpretation of \eqref{eq:ourrF} remains the same as in~\eqref{eq:rsemicl}, but now $\text{dist}(\theta)$ first finds the saddle point of the density matrix, i.e.~the largest product of the wave function $\Psi(E_n)=\mathcal{N}\exp{(-N A(\theta,\theta_n)/2)}$ and the corresponding Boltzmann factor $\exp{(-\beta E_n)}$.

We illustrate the dynamics on the Bloch sphere for a quench to $\Gamma \final=0.2$ \cite{video_thermal} at $\beta=5$ in Fig.~\ref{fig:thermal} and the corresponding LE in Fig.~\ref{fig:LE_comp}. The initial state shows a finite magnetization $m_z$ but the radius of the Bloch sphere has decreased to $s_\text{cl}\approx 0.71$. Due to the thermal fluctuations the quench is now regular and, in contrast to the $T=0$ case, shows the same features as Fig.~\ref{fig:regular}. This can be explained by the decreased spin length $s_\text{cl}<1$ which effectively renders the transverse field in the Hamiltonian more relevant compared to the $S_z^2$-term. As a result, the ground state of the final Hamiltonian is paramagnetic and the quench crosses the ferro- to paramagnetic transition in the DPT-I picture as well.

Finally, note that for high temperatures close to the equilibrium critical temperature $T_c=1/4$ the initial distribution on the Bloch sphere becomes fully determined by thermal fluctuations. Hence, $r_\text{F}$ in~\eqref{eq:ourrF} can then by replaced by the completely thermal distance measure
\begin{align}
r_\text{F}(t)=\frac{\beta s_\text{cl}^2}{8}\min_{\theta_n, \phi}\bigg\{\sin^2{\vartheta(\theta_n, \phi \,|t)}+\sin^2{\theta_n}\bigg\}\,.
\label{eq:rsimp}
\end{align}
As evidenced in Fig.~\ref{fig:LE_comp}c) this simplification already produces decent results for the quench considered in Fig.~\ref{fig:thermal}, where we are thus calculating an essentially classical return rate.

\emph{Conclusion.}--We have shown, using the example of the FC-TFIM, that for systems where the short-time dynamics is well described by MFT, the LE at zero temperature can be described to a high degree of accuracy by a semiclassical approximation. At sufficiently high temperatures, even a purely classical thermal cloud yields a qualitative reproduction of the fidelity LE that is otherwise difficult to obtain for large systems. This is remarkable since the implied approximations completely discard all dephasing, thereby prohibiting the system to relax at late times. The method also paves the way for the calculation of LEs or entanglement witnesses like Fisher information \cite{Paris2009,Marzolino2014,Marzolino2017,Braun2017} within the more general framework of the truncated Wigner approximation.

\emph{Acknowledgments.}--We thank Francesco Piazza and Wilhelm Zwerger for fruitful comments on the manuscript. This project has been supported by NIM (Nanosystems Initiative Munich).

\newpage
\bigskip
\pagebreak
\widetext
\begin{center}
\textbf{\large --- Supplemental Material ---\\ Dynamical Quantum Phase Transitions: A Geometric Picture}\\
\medskip
\text{Johannes Lang, Bernhard Frank, and Jad C. Halimeh}
\end{center}
\setcounter{equation}{0}
\setcounter{figure}{0}
\setcounter{table}{0}
\makeatletter
\renewcommand{\theequation}{S\arabic{equation}}
\renewcommand{\thefigure}{S\arabic{figure}}
\renewcommand{\bibnumfmt}[1]{[S#1]}
\renewcommand{\citenumfont}[1]{S#1}
\newcommand{\rmf}{{\rm f}}

\section{Construction of the WKB wave function}
In this supplement, we present the construction of the WKB wave function~\cite{Gottfried_book,Messiah_book} adapted to large spins~\cite{vanHemmen1986, vanHemmen2003} for our initial Hamiltonian that reads~\cite{footnote1}
\begin{align}
H_\text{i} = - \frac{1}{ 2 N }S_z^2\,.
\label{eq:S-hamiltonian}
\end{align}
At first glance, it may seem as a complete technical overkill to create a semiclassical approximation to the eigenstates of the exactly solvable $H_\text{i}$. However, the spin WKB wave functions $\Psi(q)$ can both be easily mapped onto the Bloch sphere, and, when expressed in terms of the $2S+1$ eigenstates of $S_x$ with $q \in \{-S, -S+1,\ldots,S\}$, they reproduce the correct fluctuations expressed by the expectation values of the quadratic spin operators $ \langle S^2_{\left\{x,y,z\right\}} \rangle$ \cite{footnote2}. Choosing the quantization axis along the $x$-direction transforms the Hamiltonian to $H_\text{i}=-\frac{1}{8 N } (S_+ + S_-)^2$, where $S_{+(-)}$ denotes the spin raising (lowering) operator. The stationary Schr\"odinger equation for $\Psi(q)$ with eigenenergy $E$ becomes the finite-difference equation
\begin{align}
-\frac{a^2(q)}{8N}\left[\Psi(q+2)  +2\Psi(q) + \Psi(q-2)\right] =E \Psi(q)\,,
\label{eq:S-schroedinger}
\end{align}
with boundary condition $\Psi(|q|>S)=0$. 
Here, we have omitted corrections of order $1/S$ from the exact prefactors of the raising and lowering operators and instead approximated them by $a(q)=\sqrt{S(S+1)-q^2}$. For the wave function, we use the WKB ansatz
\begin{align}
\Psi(q) = \mathcal{N}\, \eh{\text{i}\mathcal{A}(q)}\,,
\label{eq:S-ansatz}
\end{align}   
where $\mathcal N $ denotes the normalization.
Inserting this ansatz into the Schr\"odinger equation \eqref{eq:S-schroedinger} and making use of the fact that in the semiclassical limit $S \to \infty$ the argument of $\Psi(q)$ can be treated as a continuous variable, one obtains to lowest order in $1/S$
\begin{align}\label{eq:S-unmodified}
\mathcal{A}^\prime(q) =\frac{1}{2}\arccos\left(-\frac{4 N E}{a^2(q)}-1\right)\,,
\end{align}
where the prime indicates the derivative with respect to $q$.
In order to control the spin fluctuations described by the wave function with sufficient accuracy, however, it is necessary to extend the given solution by higher derivatives and thus the next-to-leading order in $1/S$, which we parametrize as
\begin{align}
\mathcal{A}^\prime(q) =\frac{1}{2}\arccos\left(-\frac{4 N E}{a^2(q)}-1+\mathcal{B}(q)\right)\,.
\label{eq:S-modified}
\end{align}
The correction term has to satisfy the condition
\begin{align}
\mathcal{B}(q)=(e^{2\text{i}\mathcal{A}^{\prime\prime}}-1)\cos{\left(2\mathcal{A}^\prime(q)\right)}\,,
\label{eq:S-unnamed}
\end{align}
which follows from the inclusion of the second derivative of $\mathcal{A}(q)$ in the continuum limit of the difference equation \eqref{eq:S-schroedinger} when acting on the ansatz \eqref{eq:S-ansatz} with modified exponent \eqref{eq:S-modified}. This expression is only a good approximation if $|\mathcal{A}^\prime(q)|\gg|\mathcal{A}^{\prime\prime}(q)|\,$, which according to \eqref{eq:S-unmodified} breaks down at the boundary of the classically allowed region. One thus has to expand $\mathcal{B}$ around the maximally forbidden spin projections $q \to S$ on the northern hemisphere (our choice for the broken $\mathbb{Z}_2$ symmetry), where successively higher derivatives are suppressed by increasing powers of $1/S$. For energies close to the ground state energy $E_0 = -S^2/(2N)$ one obtains
\begin{align}
\mathcal{B}(q) \simeq \frac{2S}{a^2(q)}\,,
\label{eq:S-Bs}
\end{align}
showing that $\mathcal{B}(q)$ is indeed a $1/S$ correction to the leading terms. With this addition,
\begin{align}
\mathcal{A}^\prime(q) = \frac{1}{2} \arccos\left(-\frac{4N E-2S}{a^2(q)}-1\right)
\label{eq:S-ouransatz}
\end{align}
is now consistent with the Schr\"odinger equation expanded up to the second derivative at all possible values of $q$.\\
The reason behind the implementation of this non-standard WKB ansatz including $\mathcal{B}(q)$ is that it changes the boundary between the classically allowed and forbidden spin orientations from $B =\sqrt{2NE+S(S+1)}\simeq \sqrt{S}$ to the far more accurate
\begin{align}
B = \sqrt{2NE+S^2} \,,
\end{align}
which vanishes for the ground state as all higher-energy spin projections can only be reached via quantum tunneling through the classically forbidden region. To obtain the final expression for $\Psi(q)$ used for the determination of the Loschmidt return rate, one has to integrate~\eqref{eq:S-ansatz} with fixed lower boundary $B$. Expanding $\mathcal{A}^\prime(q)$ around $q \simeq B$ and using the asymptotics $\arccos(z)\simeq \text{i} \sqrt{2}\sqrt{z-1}$ in the vicinity of $z \simeq 1$, results in
\begin{align}\nonumber
&\Psi(\mathcal{E},q)=\\
&\mathcal{N} \exp{\left[\frac{\sqrt{2B(S-\mathcal{E})+\mathcal{D}(q-B)}\sqrt{q-B}}{2}\left(\frac{q-B}{\mathcal{E}-S}-\frac{B}{\mathcal{D}}\right)-\frac{B^2(\mathcal{E}-S)}{\mathcal{D}^{3/2}}\ln{\frac{\sqrt{\mathcal{D}(q-B)}+\sqrt{2B(S-\mathcal{E})+\mathcal{D}(q-B)}}{\sqrt{2B(S-\mathcal{E})}}}\right]}\,.
\label{eq:S-WKBfin}
\end{align}  

\begin{figure}[htb]
    \centering
    \begin{minipage}{0.45\textwidth}
        \centering
        \includegraphics[width=\textwidth]{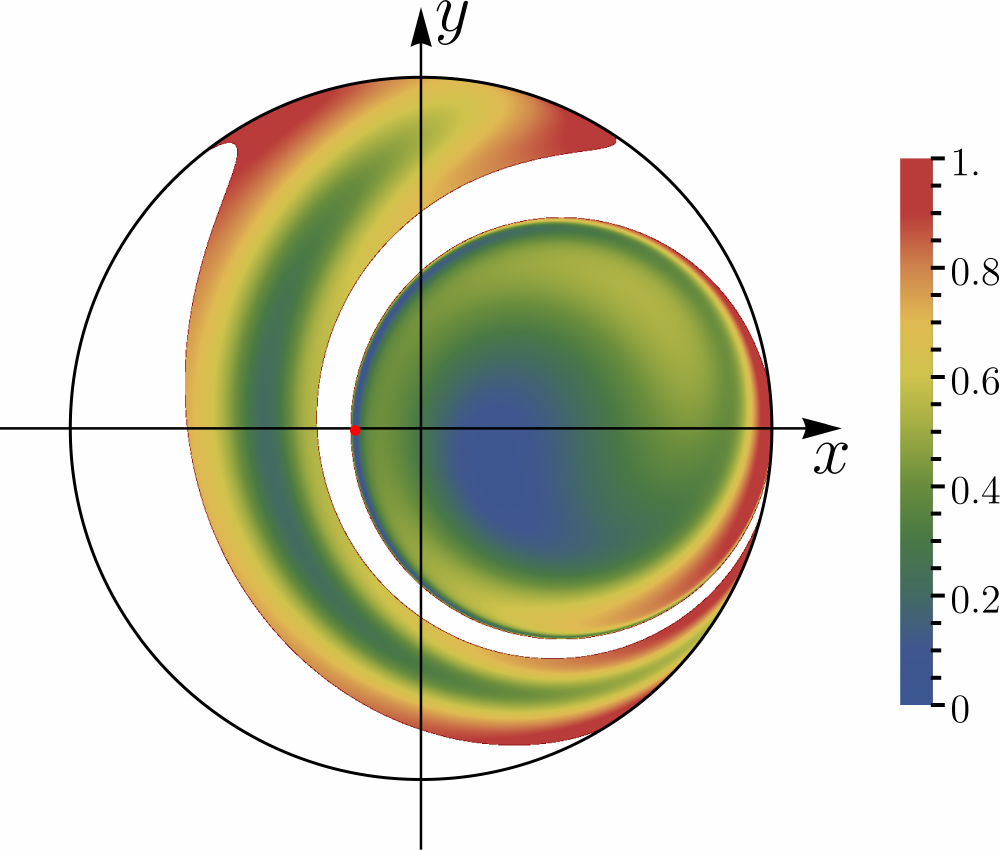} 
    \end{minipage}
     \hspace{1. cm}
    \begin{minipage}{0.45\textwidth}
        \centering
        \includegraphics[width=\textwidth]{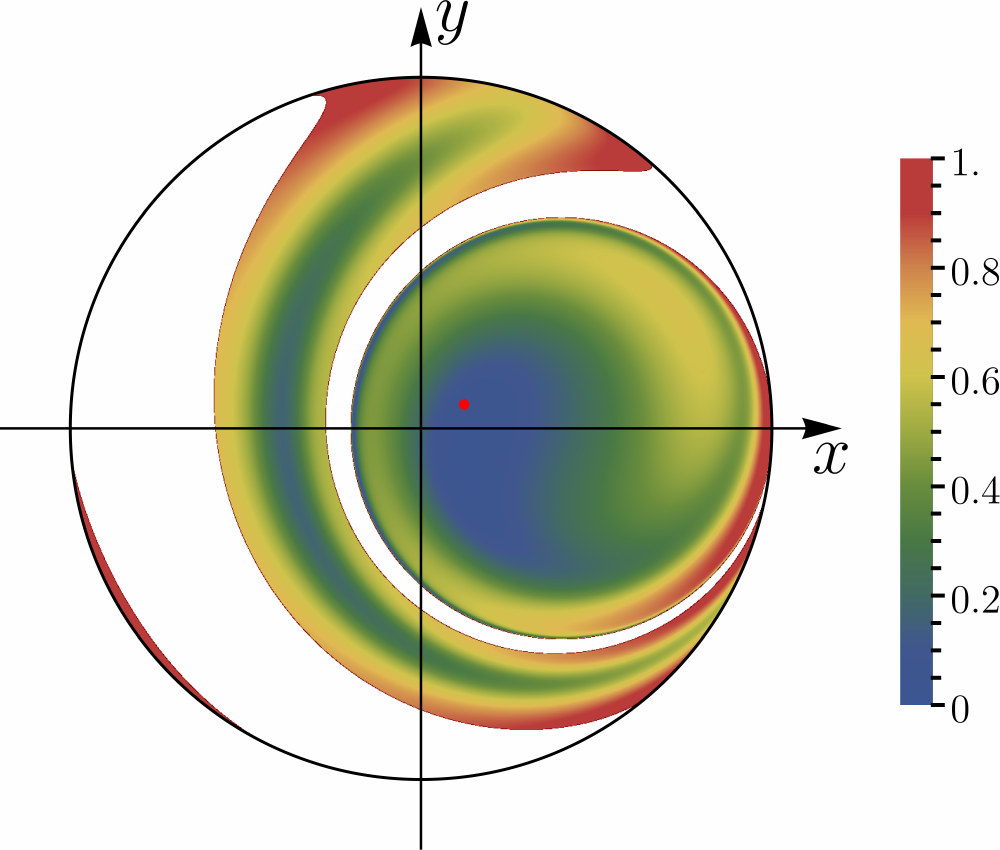} 
    \end{minipage}
\caption{Representation of the zero-temperature distance measure \eqref{eq:rsemicl} for the anomalous quench $\Gamma_\rmi=0 \to \Gamma_\rmf=0.2$ on the  Bloch sphere. The absolute minimum determines the Loschmidt  vector $\mathbf{v}_{\max}$ which is indicated by the red point. At the critical time $t_c=27.4$ it jumps discontinuously. The left panel shows the configuration at $t=t_c-1$, while the right one is taken at $t=t_c+1$. The white regions are covered by points that initially where located within the lower hemisphere. Therefore, they carry negligible semiclassical weight and have not been taken into account for the numerics.}
    \label{fig:S-dist}
\end{figure}
\begin{figure}[htb]
    \centering
    \begin{minipage}{0.45\textwidth}
        \centering
        \includegraphics[width=0.9\textwidth]{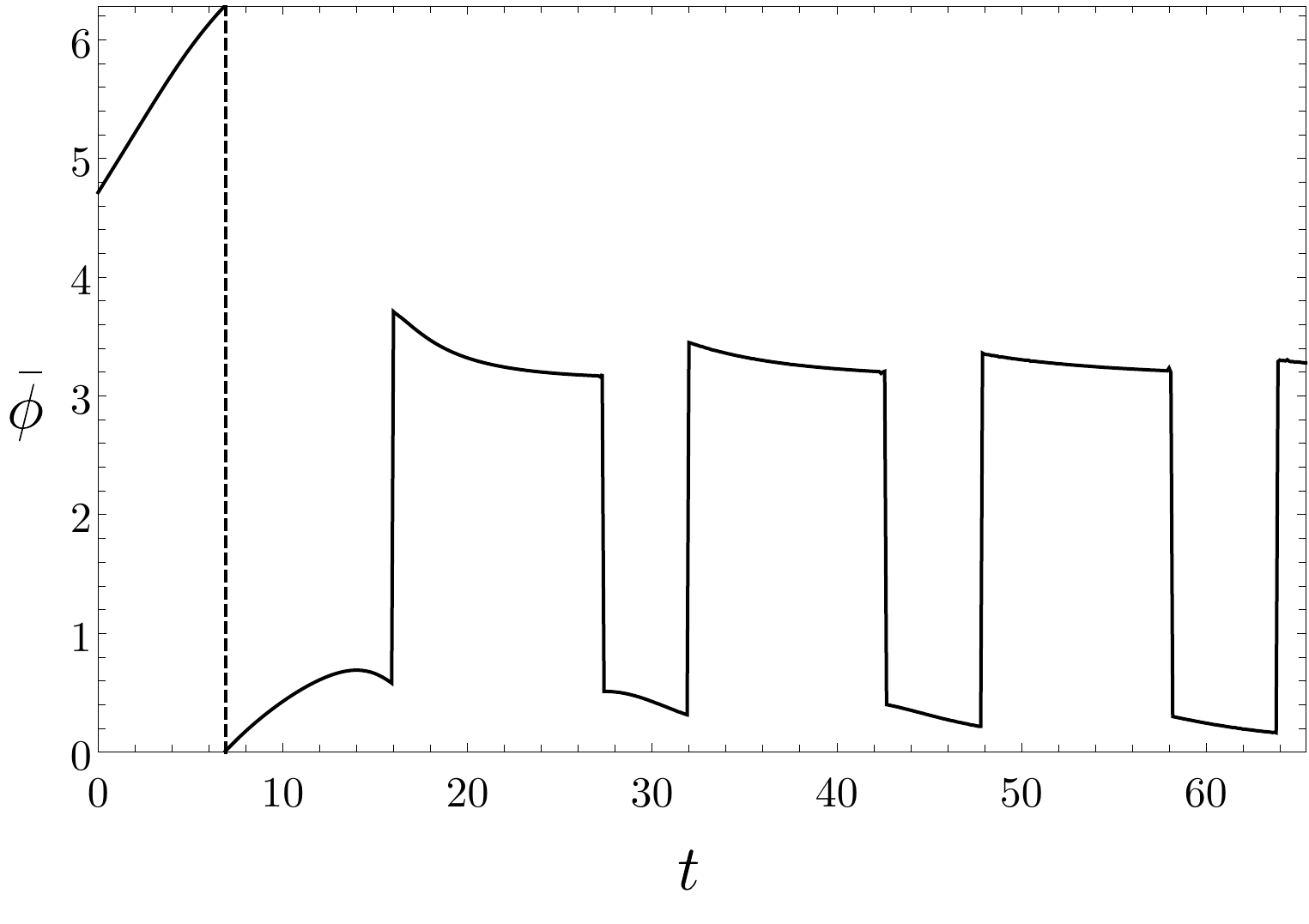} 
    \end{minipage}
    \hspace{1. cm}
    \begin{minipage}{0.45\textwidth}
        \centering
        \includegraphics[width=0.9\textwidth]{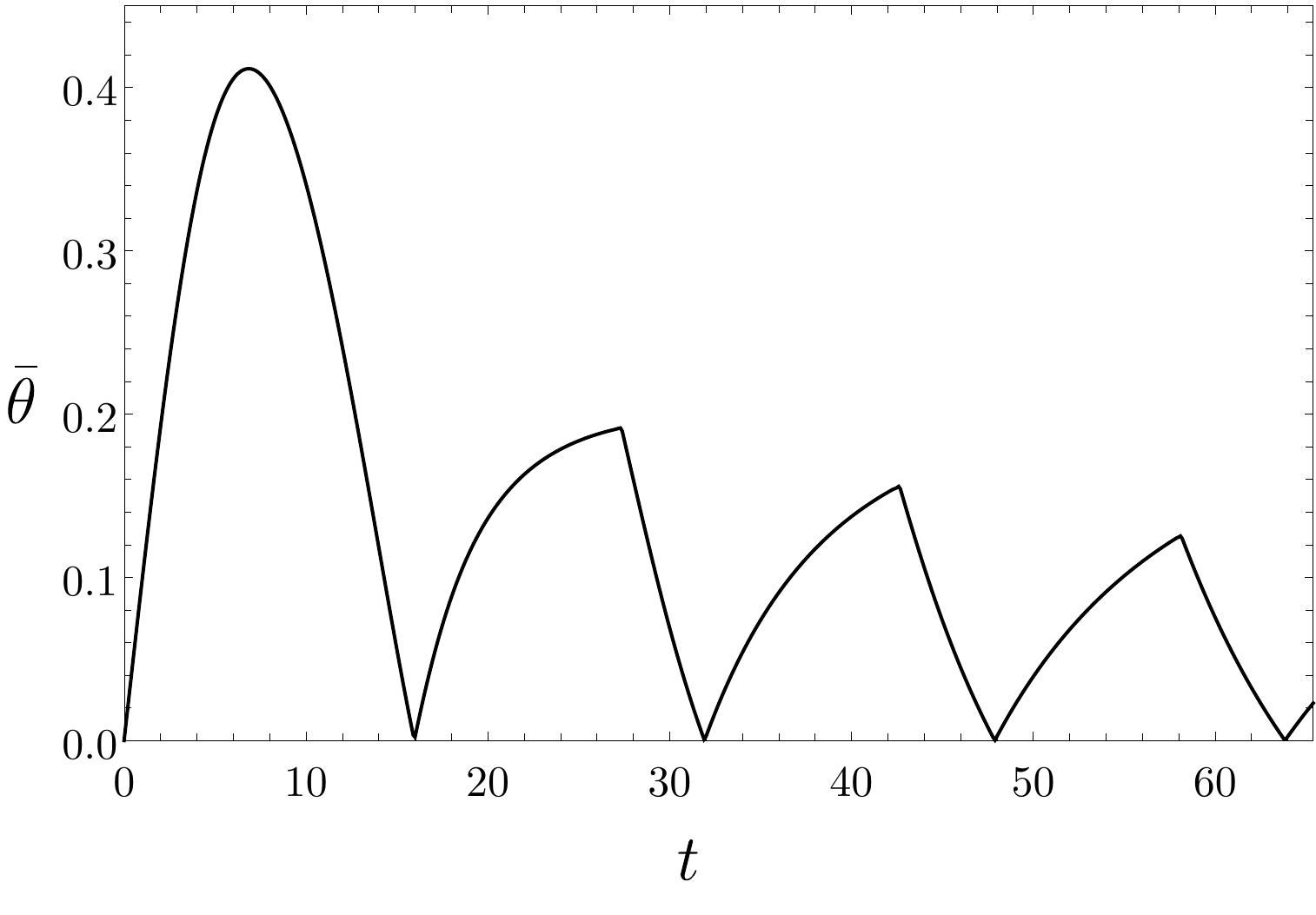} 
    \end{minipage}
    \caption{Initial angles that at time $t$ coincide with the Loschmidt vector $\mathbf{v}_\text{max}(t)=(\vartheta(\bar{\theta}, \bar{\phi} \,|t),\varphi(\bar{\theta}, \bar{\phi} \, | t))$. At the first critical time $t_c=27.4$ a jump in $\bar{\phi}$ from $3.18 \to 0.51$ occurs at fixed $\bar{\theta}$. In general, non-analyticities in $r(t)$ appear only when $\bar{\phi}$ decreases. Its sudden growths by $\pi$ in turn, are observed at times when the classical magnetization passes through the north pole, corresponding to zeros in $\bar{\theta}$ and ill-defined $\bar{\phi}$.
The dashed line in the left panel is no non-analyticity, but rather a result of the $2\pi$ periodicity of the azimuthal angle, due to which $\phi=0$ and $\phi=2\pi$ have to be identified.} 
\label{fig:S-barangles}
    \end{figure}
Here, we have introduced the abbreviations $\mathcal{E}=2NE$ and $\mathcal{D}=3\mathcal{E}+S+4 S^2\gg S$, and the normalization $\mathcal{N}$ is actually irrelevant for the determination of the Loschmidt return function. Quite importantly, $\Psi(q)$ shows the proper scaling of the spin expectations values, i.e.~for the ground state $\langle S^2_x \rangle =S/2= \langle S_y^2\rangle$ and $\langle S^2_z \rangle = S^2$ with subleading corrections.
For the geometric interpretation outlined in the main text, placing these wave functions on the Bloch sphere is simply done by substituting $q= S\sin \theta$, where $\theta$ is the polar angle on the Bloch sphere. Using the definitions for $\mathcal{E}$ and $\mathcal{D}$, $\Psi(q)$ can be further simplified into the expression~\eqref{eq:WF} given in the main text. In particular, at $T=0$, one has $B=0$ and $\mathcal{E}=-S^2 \ll -S$ yielding~\eqref{eq:WKB} in the main text. A resolution of the azimuthal angle is not necessary as the rotational symmetry of the Hamiltonian will be recovered in any eigenfunction. 

\section{Non-analytic dynamics of the Loschmidt vector}
In the main text we have argued that the Loschmidt return rate is dominated by a single trajectory on the Bloch sphere, namely the Loschmidt vector $\mathbf{v}_{\max}$, which can be obtained by minimizing a semi-classical distance measure.  
Here, we give further details on this geometric interpretation using the example of the anomalous quench from $\Gamma_\rmi = 0$ to $\Gamma_\rmf = 0.2$ at zero temperature. We show the angular distribution of the corresponding distance measure $\left(\sin^2{\vartheta(\bar\theta, \bar\phi\, |t)}+\sin^2{\bar{\theta}}\right)/2$ in Fig.~\ref{fig:S-dist}, just before and after the critical time $t_c=27.4$, when the first cusp in $r(t)$ occurs (cf. Fig.~\ref{fig:LE_comp} in the main text). At this time one of the local minima (blue regions) becomes the new global one, which causes a jump of $\mathbf{v}_{\max}$ (red dot). This discontinuous movement is directly related to the non-analyticity of $r(t)$.  

In Fig.~\ref{fig:S-barangles} we illustrate the initial orientation $(\bar\theta,\bar{\phi})$ which is found when one evolves $\mathbf{v}_{\max}(t)$ backwards in time. In agreement with the jump of the Loschmidt vector, we observe a sudden change of $(\bar\theta,\bar{\phi})$ at the critical time.


\begin{thebibliography}{6}
\bibitem[{foo({\natexlab{b}})}]{footnote2}
\bibinfo{howpublished}{The obvious, direct representation in the eigenbasis of
  $S_z$ always yields too small fluctuations due to the vanishing width of the
  WKB wave functions as $H_\text{i}$ is diagonal in that particular basis.}
\end{thebibliography}
\end{document}